\documentclass[aps, pre, showpacs, showkeys, twocolumn]{revtex4-1}

\usepackage{amsmath,amssymb,amsthm,amsfonts,latexsym}
\usepackage{bm,amsfonts, mathtools}
\usepackage{graphicx,color, wasysym}
\usepackage{textcomp}
\usepackage{amsmath,amssymb,latexsym,epsfig}
\usepackage{appendix}

\def\ulamek#1#2{\mbox{\normalfont$\frac{#1}{#2}$}}

\begin{document}

\title[The stretched exponential behavior and its underlying dynamics. The phenomenological approach]{The stretched exponential behavior and  its underlying dynamics. The phenomenological approach}

\author{K.~G\'{o}rska}
\email{katarzyna.gorska@ifj.edu.pl}
\affiliation{H. Niewodnicza\'{n}ski Institute of Nuclear Physics, Polish Academy of Sciences, Division of Theoretical Physics, ul. Eliasza-Radzikowskiego 152, PL 31-342 Krak\'{o}w, Poland}

\author{A.~Horzela}
\email{andrzej.horzela@ifj.edu.pl}
\affiliation{H. Niewodnicza\'{n}ski Institute of Nuclear Physics, Polish Academy of Sciences, Division of Theoretical Physics, ul. Eliasza-Radzikowskiego 152, PL 31-342 Krak\'{o}w, Poland}

\author{K.~A.~Penson}
\email{penson@lptl.jussieu.fr}
\affiliation{Sorbonne Universit\'{e}s, Universit\'{e} Pierre et Marie Curie (Paris 06), CNRS UMR 7600, Laboratoire de Physique Th\'{e}orique de la Mati\`{e}re Condens\'{e}e (LPTMC), Tour 13-5i\`{e}me \'{e}t., B.C. 121, 4 pl. Jussieu, F 75252 Paris Cedex 05, France}

\author{G. Dattoli}
\email{giuseppe.dattoli@enea.it}
\affiliation{ENEA - Centro Ricerche Frascati, via E. Fermi, 45, IT 00044 Frascati (Roma), Italy}

\author{G.~H.~E.~Duchamp}
\email{ghed@lipn-univ.paris13.fr}
\affiliation{Universit\'{e} Paris XIII, LIPN, Institut Galil\'{e}e, CNRS UMR 7030, 99 Av. J.-B. Clement, F 93430 Villetaneuse, France}

\medskip


 \begin{abstract}
We show that the anomalous diffusion equations with a fractional spacial derivative in the Caputo or Riesz sense are strictly related to the special convolution properties of the L\'{e}vy stable distributions which stem from the evolution properties of stretched or compressed exponential function. The formal solutions of these fractional differential equations are found by using the evolution operator method where the evolution is conceived as integral transform whose kernel is the Green function. Exact and explicit examples of the solutions are reported and studied for various fractional order of derivatives and for different initial conditions.

 \medskip

 \smallskip
 

\end{abstract}

\maketitle

\vspace*{-16pt}
 
\section{Introduction}\label{sec:1}
\setcounter{section}{1}
\setcounter{equation}{0}

In recent experiments of relaxation processes in a variety of complex materials and systems, such as supercooled liquids, spin glasses, amorphous solids, molecular systems, glass soft matter, porous and noncrystaline silicon, etc. \cite{WGotze92, JCPhillips96, CAAngell00, LCipelletti05, LPavesi96, IMihalcescu96}, the relaxation properties have been successfully fitted by stretched exponentials \cite{RKohlrausch54, GWilliams70, RSAnderssen04}
\begin{equation*}
f_{\alpha}(\ulamek{t}{\tau_{0}}) = e^{- \big(\ulamek{t}{\tau_{0}}\big)^{\alpha}}, \quad \text{for} \quad t > 0 \quad \text{and} \quad 0 < \alpha < 1,
\end{equation*}
called also the Kohlrausch-Williams-Watts functions (the KWW functions). Moreover, it has been shown, e.g. in \cite{KWeron96}, that the KWW function is related to the Cole-Cole relaxation processes which appear in the systems of disordered or anomalous structures. The importance of the KWW function for the survival propability in relation to relaxation function has been also noted and dissed in \cite{KWeron96}. The decay according to the KWW pattern is by no means restricted to common relaxation phenomena. For example, it appears in biological context, e.g. in protein folding  \cite{JSabelko99} and in $\alpha$-Helix formation in photo-switchable peptides \cite{JBredenbeck05, JAIhalainen07}, etc. The stretched exponential behavior is also used to describe the local variations of transport speed \cite{YZhang07, BDybiec08}. Exponential decay faster than the Debye law ($\alpha > 1$) is also observed. This case is called the compressed exponential, which in the theory of slow magnetization is named the Kolmogorov-Avrami-Fatuzzo relaxation \cite{HXi08, AAdanlete11, NZurauskiene14}.   

The function $f_{\alpha}(t/\tau_{0})$ is related to L\'{e}vy stable distributions \cite{HPollard46, HBergstrom52}: for $0<\alpha<1$ these are one-sided and for $1<\alpha\leq 2$ these are two-sided cases. As it is known there exists an intimate relation between the L\'{e}vy stable distributions and the anomalous diffusion; for $0<\alpha<1$ it is the sub-diffusion and for $1<\alpha\leq 2$ it is the super-diffusion. We are going to approach this connection in a novel way. The physical formulation is based on the Fokker-Planck equation with the spatial fractional derivative, usually of Riemann-Liouville, Caputo or Riesz types. The case of $\alpha = 2$ is matched with the standard diffusion equation. For $\alpha \equiv N = 3, 4, 5 \ldots$ the compressed exponential functions correspond to the partial differential equation with the space derivative of order $N$ \cite{KGorska13, EOrsinger12}. The sub- and super-diffusion equations are usually derived via the probability approach like the continuous-time random walks \cite{JKlafter80}, the master equation \cite{DBedeaux71},  the generalized master equation \cite{VMKenkre73}, and related methods \cite{ACompte96, MMagdziarz06, BDybiec10}. The intense activity in this field has apparently been initiated by Saichev and Zaslavsky in \cite{AISaichev97}. The relevant references can be traced back from two recent papers \cite{ECapelasdeOliverira14, RGarra14}. A very complete and lucid exposition of theoretical aspects of diffusion phenomena can be found in \cite{MMeerschaert12}. Nevertheless, the anomalous diffusion equations should stem from the natural evolution property which is satisfied by $f_{\alpha}(t/\tau_0)$: namely a suitably defined product of two stretched (compressed) exponential functions should give another stretched (compressed) exponential function. From the mathematical point of view, the use of the complex analysis technique underlying this condition will allow us to obtain the fractional Fokker-Planck equations. This alternative approach to a derivation of the anomalous equations and the operational method of solving them are the main objectives of this work. We believe that this will open new possibilities of looking at the stretched and compressed exponentials which have in the natural way built-in the anomalous behavior. 

The paper is organized as follows. In Sec. II we recall and comment on two relations: the first is between the stretched exponential and the one-sided L\'{e}vy stable distribution, and the second one is between the compressed exponential and two-sided L\'{e}vy stable distribution. We use the well-known unique relation between the one- and two-variables L\'{e}vy stable distributions whose evolution property will be studied in the paper. In Sec. II we derive two types of integral evolution equations related to two various behaviors of $f_{\alpha}(t/\tau_0)$. The differential version of these equations will be found in Sec. III. In connection with the used L\'{e}vy stable distribution they will contain the fractional derivative in the Caputo or Riesz sense. The solution of these differential equations in terms of the Green function are given in Sec. IV. Sec. V is devoted to finding the solution of these equations with the help of operational methods based on the integral transforms whose kernels are given by the appropriate L\'{e}vy stable distributions. Sec. VI contains the explicit examples of solutions of the equations. We conclude the paper in Sec. VII.

\section{L\'{e}vy stable distributions}\label{sec:2}

\setcounter{section}{2}
\setcounter{equation}{0}

In a discrete setting and from a phenomenological point of view the KWW functions for $0 < \alpha < 1$ can be treated as a sum of weight function of exponential decays ($f_{\alpha}(t/\tau_0)$ for $\alpha =1$) \cite{GDattoli14}, namely 
\begin{equation}\label{eq1}
f_{\alpha}(\ulamek{t}{\tau_0}) = \sum_{j} e^{- \frac{t}{\tau_0} u_{j}} g_{\alpha}(u_{j}) \Delta u_{j}, 
\end{equation}
where $g_{\alpha}(u_{j})$ is a probability density, i.e. $g_{\alpha}(u_{j}) > 0$ and $\sum_{j} g_{\alpha}(u_{j})\Delta u_{j}~=~1$, which describes the structure of the sample where relaxation process is governed by the stretched exponential function. In the limit of large number of elements in the sum and for the infinitesimally small changes $\Delta u_{i}$, Eq. (\ref{eq1}) goes over to the Laplace transform \cite{GDattoli14, KWeron91}
\begin{align}\label{eq2}
f_{\alpha}(\ulamek{t}{\tau_0}) = \mathcal{L}\left[g_{\alpha}(u); \ulamek{t}{\tau_0}\right] & = \int_0^\infty e^{- \frac{t}{\tau_0} u} g_{\alpha}(u) du \\
& = \int_0^\infty e^{- x} g_{\alpha}((\ulamek{t}{\tau_0})^{\alpha}, x) dx. \label{eq3}
\end{align}
In Eq. (\ref{eq2}) the variable of integration is changed to $u~=~\tau_0 x/t$ and it gives the relation \cite{VMZolotarev83, KGorska12, KAPenson16}:
\begin{equation}\label{eq4}
\ulamek{\tau_0}{t} g_{\alpha} (\ulamek{\tau_0}{t} x) \equiv g_{\alpha}((\ulamek{t}{\tau_0})^{\alpha}, x). 
\end{equation}
Systematic treatment of relations of type Eq. \eqref{eq4} can be rephrased in terms of subordination laws, consequences of the properties of Mellin transform \cite{BDybiec10, KAPenson16, FMainardi03}. Eq. (\ref{eq4}) represents the one-to-one correspondence between the \textit{one-} and \textit{two-} variables L\'{e}vy stable distributions. It leads to the evolution equation with the fractional derivative which will be studied below. In this sense Eq. (\ref{eq4}) is crucial for our considerations.

In order to limit the proliferation of notation we shall employ the same symbol to both sides of Eq. (\ref{eq4}), and later on, to Eq. (\ref{eq10}). Their meaning will be clear from the context and should not lead to confusion.

The probability density $g_{\alpha}((\ulamek{t}{\tau_0})^{\alpha}, x)$, via the inverse Laplace transform of Eq. (\ref{eq2}) and Eq. (\ref{eq4}), is uniquely given by one-sided L\'{e}vy stable distributions \cite{HPollard46}:
\begin{align}\label{eq5}
\begin{split}
g_{\alpha}((\ulamek{t}{\tau_0})^{\alpha}, x) & = \mathcal{L}^{-1}\{\exp[-\big(\ulamek{t}{\tau_0}\xi\big)^{\alpha}]; x\} \\
& = \frac{1}{2\pi i} \int_{L} e^{x \xi} e^{-\big(\ulamek{t}{\tau_0}\xi\big)^{\alpha}} d\xi,
\end{split}
\end{align}
$ x > 0$, and $g_{\alpha}((\ulamek{t}{\tau_0})^{\alpha}, x) = 0$ for $x\leq 0$. We remark that the Laplace transform $f_{L}(p)$ of the function $f(u)$ is defined as follows \cite{ INSneddon74}:
\begin{equation*}
f_{L}(p) = \mathcal{L}[f(u); p] = \int_{0}^{\infty} e^{- u p} f(u) du,
\end{equation*}
for $p\in\mathbb{C}$, whereas the inverse Laplace transform is given by
\begin{equation*}
f(u) = \mathcal{L}^{-1}[f_{L}(p); u] = \frac{1}{2\pi i} \int_{L} e^{u p} f_{L}(p) dp,
\end{equation*}
where inside the integration contour $L$ the function $f_{L}(p)$ is analytic. Eq. (\ref{eq5}) goes to the Dirac $\delta$ function, namely $\delta(u)$, in the limit of $t\to 0$, see \cite[Eq. (17.13.94) on p. 1115]{ISGradsteyn07}. The function $g_{\alpha}((\ulamek{t}{\tau_0})^{\alpha}, x)$ has the essential singularity at $x = 0$ \cite[Eq. (4)]{JMikusinski59} and the ``heavy'' long tail as $x \to \infty$ \cite[Eq. (5)]{JMikusinski59}. Taking in Eq. (\ref{eq5}) the integral contour $L$ defined in \cite{HPollard46} and using Eq. (\ref{eq4}), we get 
\begin{equation}\label{eq6}
g_{\alpha}((\ulamek{t}{\tau_0})^{\alpha}, x) = {\rm Im}\left\{\int_{0}^{\infty} e^{- x \xi} e^{-(\frac{t}{\tau_0} \xi)^{\alpha}e^{-i \pi \alpha}} \frac{d\xi}{\pi}\right\},
\end{equation}
which for rational $0 < \alpha = l/k < 1$ is equal to
\begin{align}\label{eq7}
\begin{split}
g_{\frac{l}{k}}((\ulamek{t}{\tau_0})^{\frac{l}{k}}, x) & = \sum_{j=1}^{k-1} \frac{(-1)^{j} x^{-1-j\frac{l}{k}}}{j!\, \Gamma(-\frac{l}{k}j)} \left(\!\frac{t}{\tau_0}\!\right)^{j\frac{l}{k}}  \\
&\times{_{l+1}F_{k}}\left({1, \Delta(l, 1 + j \frac{l}{k}) \atop \Delta(k, 1+j)}\Big\vert z\right),
\end{split}
\end{align}
with $z= (-1)^{k-l} (t l)^l/[k^k (\tau_0 x)^l]$, see \cite{KAPenson10}. The generalized hypergeometric function is denoted as ${_{n}F_{m}}\left({(\alpha_{n}) \atop (\beta_{m})} \Big\vert x\right)$, where $(\alpha_{n})$ is the ``upper'' list of parameters and $(\beta_{m})$ is the ``lower'' list \cite{APPrudnikov_v3}. The symbol $\Delta(n, a)$ is equal to a list of $n$ elements $\frac{a}{n}, \frac{a+1}{n}, \ldots, \frac{a+n-1}{n}$. The basic example of $g_{l/k}((\ulamek{t}{\tau_0})^{l/k}, x)$ is the so-called L\'{e}vy-Smirnov distribution $g_{1/2}((\ulamek{t}{\tau_0})^{1/2}, x) = \sqrt{t/\tau_0}\exp[-(t/\tau_0)/(4x)]/(2\sqrt{\pi} x^{3/2})$. The other explicit and exact examples of $g_{\alpha}((\ulamek{t}{\tau_0})^{l/k}, x)$ can be obtained by using Eq. (\ref{eq4}) and further examples are quoted in \cite{KAPenson10}. 

The compressed exponential behavior can be written as $f_{\alpha}(t/\tau_{0}) = \ulamek{1}{2} f_{\alpha}(|t/\tau_0|)$, which for $1 < \alpha < 2$ is related via the Fourier transform \cite{INSneddon74} to $\tilde{g}_{\alpha}(u)$, the two-sided L\'{e}vy distribution,
\begin{align}\label{eq8}
f_{\alpha}(\ulamek{t}{\tau_0}) = \ulamek{1}{2} f_{\alpha}(|\ulamek{t}{\tau_0}|) &= \mathcal{F}\left[\ulamek{1}{2} \tilde{g}_{\alpha}(u); \ulamek{t}{\tau_0}\right] \nonumber\\
& = \int_{-\infty}^{\infty} e^{i \ulamek{t}{\tau_0} u} \, \ulamek{1}{2} \tilde{g}_{\alpha}(u) du \\ \label{eq9}
& = \int_{-\infty}^{\infty} e^{i x}  \, \ulamek{1}{2} \tilde{g}_{\alpha}((\ulamek{t}{\tau_0})^{\alpha}, x) dx,
\end{align}
where, similarly to Eqs. (\ref{eq2}) and (\ref{eq3}), we use the new variable of integration $x$. Let us recall that the Fourier transform $g_{F}(\omega)$ of the function $g(x)$ is given by
\begin{equation*}
g_{F}(\omega) = \mathcal{F}[g(x); \omega] = \int_{-\infty}^{\infty} e^{-i \omega x} g(x) dx,
\end{equation*}
where 
\begin{equation*}
g(x) = \mathcal{F}^{-1}[g_{F}(\omega); x] = \frac{1}{2\pi} \int_{-\infty}^{\infty} e^{i \omega x} g_{F}(\omega),
\end{equation*}
which is the inverse Fourier transform. In analogy to Eq. (\ref{eq4}) we get the unique scaling relation
\begin{equation}\label{eq10}
\ulamek{\tau_0}{t} \tilde{g}_{\alpha} (\ulamek{\tau_0}{t} x) \equiv \tilde{g}_{\alpha}((\ulamek{t}{\tau_0})^{\alpha}, x), 
\end{equation}
whose evolution property will be studied in this work. The inversion of  the Fourier transform given in  Eq. (\ref{eq8}) and the use of property (\ref{eq10}) leads to
\begin{align}\label{eq11}
\begin{split}
\tilde{g}_{\alpha}((\ulamek{t}{\tau_0})^{\alpha}, x) & = \mathcal{F}^{-1}\{\exp[-\big(\!\ulamek{t}{\tau_0}\!\big)^{\alpha}|\xi|^{\alpha}]; x\} \\
& = \int_{-\infty}^{\infty} e^{- i x \xi} e^{-\big(\!\ulamek{t}{\tau_0}\!\big)^{\alpha}|\xi|^{\alpha}} \frac{d\xi}{2 \pi} 
\end{split}
\end{align}
for $x\in\mathbb{R}$ and $t > 0$, which is the symmetric two-sided L\'{e}vy stable distribution, see Eq. (8) for $\beta = 0$ of \cite{HBergstrom52} applied for (\ref{eq10}). Observe that Eq. (\ref{eq11}) defines also the symmetric L\'{e}vy stable distribution for $0 < \alpha < 1$ \cite{HBergstrom52, KGorska11}. In the limit of vanishing $t$ the two-sided L\'{e}vy stable distribution goes to the Dirac $\delta$ function, i.e. $\tilde{g}_{\alpha}((\ulamek{t}{\tau_0})^{\alpha}, x) \to \delta(x)$ for $t\to 0$. Eq. (\ref{eq11}) expressed as the finite sum of the generalized hypergeometric function is given in \cite[Eqs. (4) and (5) for $l=2p$, $k=2q$, and $r=p$]{KGorska11}, namely
\begin{align}
\tilde{g}_{\frac{p}{q}}((\ulamek{t}{\tau_0})^{\frac{p}{q}}, x) & = \sum_{j=1}^{M-1} \frac{(-1)^{1+j}}{\pi\, j!}\, \Gamma\left(1 + \frac{m}{M} j\right) \frac{(t/\tau_0)^{\mp \frac{2p}{M} j}}{x^{1\mp\frac{2 p}{M}j}} \nonumber \\
& \times {_{m+1}F_{M}}\left({1, \Delta(m, 1 + \frac{m}{M}j) \atop \Delta(M, 1+j)} \Big\vert \tilde{z}\right), \label{eq12}
\end{align}
where $\tilde{z} = (-1)^{p-M}[t/(\tau_0 x)]^{\mp 2 p}\, m^m/M^M$, $m = \min(2p, 2q)$, $M = \max(2p, 2q)$. The lower sign in the powers of $t/\tau_{0}$ and $x$ is for $p < q$, whereas the upper sign is for $p > q$. For $\alpha = 2$ we get the Gauss (normal) distribution, $\tilde{g}_{2}((t/\tau_0)^2; x) = \tau_0\exp[-(x\tau_0)^2/(2 t)^2]/(4 \sqrt{\pi} t)$. For more examples see \cite{KGorska11}.

From the very natural evolution assumption written for the stretched (compressed) exponential function for $t_0 \leq t_1 \leq t_2$, it appears that an appropriately defined composition of two stretched (compressed) exponentials should give another stretched (compressed) exponential. It means that the appropriate convolution of the stable distributions of two variables gives another stable distribution of two variables: (i) for the one-sided L\'{e}vy stable distribution of two variables we have 
\begin{equation}\label{eq13}
g_{\alpha}((\ulamek{t_2-t_0}{\tau_0})^{\alpha}, x) \!=\! \int_{0}^{x}\!\!\! g_{\alpha}((\ulamek{t_2-t_1}{\tau_0})^{\alpha}, y) g_{\alpha}((\ulamek{t_1-t_0}{\tau_0})^{\alpha}, x-y) dy,
\end{equation}
whereas (ii) for the two-sided L\'{e}vy stable distribution of two variables we get 
\begin{equation}\label{eq14}
\tilde{g}_{\alpha}((\ulamek{t_2-t_0}{\tau_0})^{\alpha}, x) \!=\! \int_{-\infty}^{\infty}\!\!\! \tilde{g}_{\alpha}((\ulamek{t_2-t_1}{\tau_0})^{\alpha}, y) \tilde{g}_{\alpha}((\ulamek{t_1-t_0}{\tau_0})^{\alpha}, x-y) dy.
\end{equation}
The proofs of Eqs. (\ref{eq13}) and (\ref{eq14}) are given in Appendix A. Eqs. (\ref{eq13}) and (\ref{eq14}) are the \textit{new} types of Laplace and Fourier convolutions, respectively, where both of the arguments of functions, i.e. $t$ \textit{and} $x$, vary. We point out that Eqs. (\ref{eq13}) and (\ref{eq14}) differ from the standard Laplace and Fourier convolutions of the \textit{one} variable L\'{e}vy stable distributions which lead to the stable laws \cite{VMZolotarev83, WFeller71}. For the stretched exponential the stable law has the form
\begin{equation*}
e^{-(b_1 p)^{\alpha}} e^{-(b_2 p)^{\alpha}} = e^{-(b p)^{\alpha}} \quad \text{for} \quad b_{1}^{\alpha} + b_{2}^{\alpha} = b^{\alpha}.
\end{equation*}

Eqs. (\ref{eq13}) and (\ref{eq14}) are the integral forms of the evolution equations whose differential forms will be found in the next section.

The properties inherent in Eqs. (\ref{eq13}) and (\ref{eq14}) naturally illustrate the transitivity of evolution expressed in two-variable version of $g_{\alpha}$'s and $\tilde{g}_{\alpha}$'s. Apparently, these extensions were not employed in the current context before.

\section{Fractional differential equation related to L\'{e}vy stable distributions}\label{sec:3}
\setcounter{section}{3}
\setcounter{equation}{0}

\subsection{The differential form of Eq. (\ref{eq13})}

We start by rephrasing Eq. (\ref{eq13}) for $(t_2 - t_1)/\tau_0 = T$ and infinitesimally small $(t_1 - t_0)/\tau_0 = \Delta T$. The r.h.s of Eq. (\ref{eq13}) can be estimated by taking the asymptotics of $g_{\alpha}((\Delta T)^{\alpha}, x-y)$ for $\Delta T \ll 1$. Using Eq. (\ref{eq6}), where instead of $\exp[-(\Delta T \xi)^{\alpha} e^{-i\pi\alpha}]$ we take its first two terms of asymptotic expansion, i.e. $1 - (\Delta T\, \xi)^{\alpha} e^{-i\pi\alpha}$, we obtain
\begin{align}\label{eq15}
& g_{\alpha}((\Delta T)^{\alpha}, x-y) \simeq {\rm Im}\left\{\int_0^\infty e^{-(x-y)\xi}\frac{d\xi}{\pi}\right\} \\ \label{eq16}
& \quad - {\rm Im}\left\{(\Delta T)^{\alpha}  e^{-i\pi\alpha} \int_0^\infty e^{-(x-y)\xi}\xi^{\alpha}\frac{d\xi}{\pi} \right\} \\ \label{eq17}
& \quad = {\rm Im}\left\{\frac{1}{\pi(x-y)} - (\Delta T)^{\alpha} e^{-i\pi\alpha} \frac{\Gamma(1+\alpha)}{\pi(x-y)^{1+\alpha}}\right\}. 
\end{align}
In Eqs. (\ref{eq15}) and (\ref{eq16}) we employed \cite[Eq. (2.3.3.1) on p. 322]{APPrudnikov_v1}. Substitution of Eq. (\ref{eq17}) into Eq. (\ref{eq13}) written for $(t_2 - t_1)/\tau_0 = T$ and $(t_1 - t_0)/\tau_0 = \Delta T$ gives
\begin{align}\label{eq18}
&\text{RHS of Eq. } (\ref{eq13}) = {\rm Im}\left\{ \int_0^x \frac{g_{\alpha}(T^{\alpha}, y)}{x-y} \frac{dy}{\pi} \right\} \\
& \qquad - {\rm Im}\left\{\frac{(\Delta T)^{\alpha}}{e^{i\pi\alpha}} \Gamma(1+\alpha) \int_0^x \frac{g_{\alpha}(T^{\alpha}, y)}{(x-y)^{1+\alpha}} \frac{dy}{\pi}\right\}. \label{eq19}
\end{align}

The integral in Eq. (\ref{eq18}) can be calculated by inserting Eq. (\ref{eq6}) into it, and then changing the order of integration. That leads to
\begin{align}\label{eq20}
& {\rm Im}\left\{ \int_0^x \frac{g_{\alpha}(T^{\alpha}, y)}{x-y} \frac{dy}{\pi} \right\} = {\rm Im}\left\{\int_0^\infty e^{-(T\xi)^{\alpha}\exp(-i\alpha\pi)} \right\} \nonumber\\
& \qquad \times\left. {\rm Im}\left[\int_0^x \frac{e^{-y\xi}}{x-y} dy\right]\frac{d\xi}{\pi^2}\right\} \\
& \qquad = {\rm Im}\left\{\int_0^\infty e^{-x\xi} e^{-(T\xi)^{\alpha}\exp(-i\alpha\pi)} \frac{d\xi}{\pi}\right\} \label{eq21} \\
& \qquad = g_{\alpha}(T^{\alpha}, x). \nonumber 
\end{align}
The second integral in Eq. (\ref{eq20}) is calculated in Appendix B and it is equal to $\pi e^{-x\xi}$, see Eq. (\ref{eq60}). In Eq. (\ref{eq21}) we employ the definition given in Eq. (\ref{eq6}).

Next, we calculate the integral in Eq. (\ref{eq19}) as follows: we apply $(x-y)^{-\alpha-1} = \frac{1}{\alpha} \frac{d}{d y}(x-y)^{-\alpha}$, integrate Eq. (\ref{eq19}) by parts and use the definition of the one-sided L\'{e}vy stable distribution in Eq. (\ref{eq6}). Thus, 
\begin{align}
& {\rm Im}\left\{\frac{\Gamma(1+\alpha)}{e^{i\pi\alpha}} \int_0^x \frac{g_{\alpha}(T^{\alpha}, y)}{(x-y)^{1+\alpha}} \frac{dy}{\pi}\right\}  \nonumber\\
& \qquad\qquad = \Gamma(\alpha)\, {\rm Im}\left[\lim_{y\to x} \frac{g_{\alpha}(T^{\alpha}, y)}{(y-x)^{\alpha}} \right] \label{eq22} \\
&\qquad\qquad - {\rm Im}\left[\frac{\Gamma(\alpha)}{e^{i\pi\alpha}} \int_0^x \frac{g'_{\alpha}(T^{\alpha}, y)}{(x-y)^{\alpha}} \frac{dy}{\pi}\right],  \label{eq23}
\end{align}
where $g'_{\alpha}(T^{\alpha}, y) = \frac{d}{dy} g_{\alpha}(T^{\alpha}, y)$. From the theory of residues \cite{FWByron92} the limit in Eq. (\ref{eq22}) is equal to
\begin{equation}\label{eq24}
\lim_{y\to x} \frac{g_{\alpha}(T^{\alpha}, y)}{(y-x)^{\alpha}} = \frac{1}{2\pi i} \oint_{S_{\delta}} \frac{g_{\alpha}(T^{\alpha}, z)}{(z-x)^{\alpha}} dz.
\end{equation}
The contour $S_{\delta}$ is a circle around $x$ with the radius $\delta$ ($\delta\ll 1$) in a counterclockwise manner. Inserting Eq. (\ref{eq6}) into Eq. (\ref{eq24}) and changing the order of integration we have
\begin{align*}
{\rm Im}\left[\lim_{y\to x} \frac{g_{\alpha}(T^{\alpha}, y)}{(y-x)^{\alpha}}\right] & = {\rm Im}\left\{ \int_0^\infty e^{-(T \xi)^{\alpha} \exp(-i\pi\alpha)} \right. \\ 
& \times \left. {\rm Im} \left[\frac{1}{2\pi i}  \oint_{S_{\delta}} \frac{e^{-z\xi}}{(z-x)^{\alpha}} dz\right] \frac{d\xi}{\pi}\right\}.
\end{align*}
Setting $z-x = \delta e^{i\varphi}$ and then taking the limit of small $\delta$ we obtain that the integral over closed contour $S_{\delta}$ has only the real value which is equal to $e^{-x\xi}$. That leads to vanishing of the integral and consequently to vanishing Eq. (\ref{eq22}). Consequently, Eq. (\ref{eq23}) is equal to
\begin{equation}\label{eq25}
{\rm Im}\left[\frac{\Gamma(\alpha)}{e^{i\pi\alpha}} \int_0^x \frac{g'_{\alpha}(T^{\alpha} y)}{(x-y)^{\alpha}} \frac{dy}{\pi}\right] = \int_{0}^{x} \frac{g'_{\alpha}(T^{\alpha}, y)}{(x-y)^{\alpha}} \frac{dy}{\Gamma(1-\alpha)},
\end{equation}
which follows from \cite[Eq. (8.334.3)]{ISGradsteyn07} and the fact that the integral in Eq. (\ref{eq25}) is real. That can be shown by calculating the derivative of $g_{\alpha}(T^{\alpha}, y)$ given in Eq. (\ref{eq6}), inserting it in Eq. (\ref{eq23}), changing the order of integration, and using \cite[Eq. (3.381.1) on p. 346]{ISGradsteyn07}. 

Expressing the l.h.s. of Eq. (\ref{eq13}) written for $(t_2~-~t_1)/\tau_0 = T$ and infinitesimally small $(t_1 - t_0)/\tau_0 = \Delta T$ into the Taylor series up to the second term we get
\begin{align}\label{eq26}
& g_{\alpha}((T+\Delta T)^{\alpha}, x) \simeq g_{\alpha}(T^{\alpha}, x) +  \partial_{T^{\alpha}} g_{\alpha}(T^{\alpha}, x) \nonumber\\
& \qquad \times [(T + \Delta T)^{\alpha} - T^{\alpha}] \\ 
& \qquad \simeq g_{\alpha}(T^{\alpha}, x) +  (\Delta T)^{\alpha} \partial_{T^{\alpha}} g_{\alpha}(T^{\alpha}, x), \label{eq27}
\end{align}
where in Eq. (\ref{eq26}) we employed the asymptotics of the square bracket, namely $[(T + \Delta T)^{\alpha} - T^{\alpha}] \simeq (\Delta T)^{\alpha}$. 

Collecting together all the contributions we obtain that the right hand side of Eq. (\ref{eq13}) for $(t_2-t_1)/\tau_0 = T$ and $(t_1-t_0)/\tau_0 = \Delta T$, where $\Delta T \ll1$, has the form specified below:
\begin{equation}\label{eq28}
\partial_{T^{\alpha}} g_{\alpha}(T^{\alpha}, x) = -\, {^{C}\!\partial_{x}^{\alpha}} g_{\alpha}(T^{\alpha}, x), \quad 0 < \alpha < 1.
\end{equation}
In Eq. (\ref{eq28}) ${^{C}\!\partial_{x}^{\alpha}} g_{\alpha}(T^{\alpha}, x)$ denotes the so-called fractional derivative in the Caputo sense \cite{IPodlubny99, SGSamko93, Mainardi1} 
\begin{equation*}
{^{C}\!\partial_{x}^{\alpha}} f(x) = \frac{1}{\Gamma(1-\alpha)} \int_{0}^{x} \frac{f'(y) dy}{(x-y)^{\alpha}},
\end{equation*}
where $0 < \alpha < 1$ and $f'(y) = \frac{d}{d y} f(y)$.

\subsection{The differential form of Eq. (\ref{eq14})}

Let us now find the differential form of  Eq. (\ref{eq14}) for $(t_2 - t_1)/\tau_0 = T$ and $(t_1 - t_0)/\tau_0 = \Delta T$ such that $\Delta T \ll 1$. Similarly to the approach presented in the previous subsection we will estimate the r.h.s. of Eq. (\ref{eq14}) for infinitesimally small values of $\Delta T$. Employing Eq. (\ref{eq11}) adjusted for $\tilde{g}_{\alpha}((\Delta T)^{\alpha}, x-y)$ and taking the asymptotic expansion of $\exp[-(\Delta T)^{\alpha} |\xi|^{\alpha}]$ up to the second term, we obtain
\begin{align}\label{eq29}
& \tilde{g}_{\alpha}((\Delta T)^{\alpha}, x-y) \simeq \int_{-\infty}^{\infty} e^{-i(x-y)\xi} [1 - (\Delta T)^{\alpha} |\xi|^{\alpha}] \frac{d\xi}{2\pi} \nonumber \\
& \qquad = \delta(x-y) - \frac{(\Delta T)^{\alpha}}{2} \frac{|x-y|^{-1-\alpha}}{\cos(\ulamek{\alpha\pi}{2}) \Gamma(-\alpha)}.
\end{align}
In Eq. (\ref{eq29}) we have used formulas (1.1) on p. 102 and (1.11) on p. 103 of \cite{YABrychkov77}. Inserting Eq. (\ref{eq29}) into the r.h.s. of Eq. (\ref{eq14}) gives 
\begin{align}\label{eq30}
\begin{split}
\text{RHS of Eq. }(\ref{eq14}) & = \tilde{g}_{\alpha}(T^{\alpha}, x) - \frac{(\Delta T)^{\alpha}}{2\Gamma(-\alpha) \cos(\ulamek{\pi \alpha}{2})} \\
& \times \int_{-\infty}^{\infty} \frac{\tilde{g}_{\alpha}(T^{\alpha}, y)}{|x-y|^{1+\alpha}} dy.
\end{split}
\end{align}
The integral in Eq. (\ref{eq30}) is related to the fractional derivative in the Riesz sense \cite{RGorenflo98, SGSamko93}:
\begin{equation}\label{eq31}
{^{R}\partial_{|x|}^{\alpha}} f(x) = -\frac{1}{2 \Gamma(-\alpha) \cos(\ulamek{\alpha\pi}{2})} \int_{-\infty}^{\infty}\frac{f(y)\, dy}{|x-y|^{1+\alpha}}
\end{equation}
for $0 < \alpha < 1$. 

The first two terms of the Taylor series of the l.h.s. of Eq. (\ref{eq14}) read
\begin{equation}\label{eq32}
\tilde{g}_{\alpha}((T +\Delta T)^{\alpha}, x)  \simeq \tilde{g}_{\alpha}(T^{\alpha}, x) + (\Delta T)^{\alpha} \, \partial_{T^{\alpha}} \tilde{g}_{\alpha}(T^{\alpha}, x),
\end{equation}
which is in analogy to Eq. (\ref{eq27}). Substituting Eqs. (\ref{eq32}), (\ref{eq30}) and (\ref{eq31}) into Eq. (\ref{eq14}) with $(t_2-t_1)/\tau_0 = T$ and infinitesimally small $(t_1 - t_0)/\tau_0 = \Delta T$ we obtain
\begin{equation}\label{eq33}
\partial_{T^{\alpha}} \tilde{g}_{\alpha}(T^{\alpha}, x) = {^{R}\partial_{|x|}^{\alpha}} \tilde{g}_{\alpha}(T^{\alpha}, x)
\end{equation}
for $1 < \alpha \leq 2$, see also \cite{Mainardi1, Mainardi}. 

\section{The Green function method}\label{sec:4}
\setcounter{section}{4}
\setcounter{equation}{0}

We recall the fundamental facts about the propagator and the Green function, compare \cite{FWByron92}. We consider the differential equation in the form 
\begin{equation}\label{eq34}
\partial_y h(y, x) = \hat{O} h(y, x), 
\end{equation}
where an operator $\hat{O} = \hat{O}(x, \frac{d}{d x})$ is independent on $y$. Its formal solution for $y > y'$ can be given by
\begin{equation*}
h(y, x) = \int K(y, x; y', x') h(y', x') dx'.
\end{equation*}
The function $K(y, x; y', x')$ is a propagator related to Eq. (\ref{eq34}) which is an response of the system on the given initial (or boundary) conditions which are the Dirac $\delta$ functions. The propagator $K$ has the following properties: (i) $K(y, x; y, x') = \delta(x-x')$; (ii) $(\partial_y - \hat{O}) K(y, x; y', x') = 0$; and (iii) for $y_0 \leq y_1 \leq y_2$ the propagator satisfies $K(y_2, x_2; y_0, x_0) = \int K(y_2, x_2; y_1, x_1) K(y_1, x_1; y_0, x_0) dx_1$, which is the Laplace convolution for the one-sided L\'{e}vy stable distribution and the Fourier convolution for the two-sided L\'{e}vy distribution. 

$K(y, x; y', x')$ can be used to define the Green function $G(y, x; y', x')$, which fulfills 
\begin{equation}\label{eq35}
(\partial_{y} - \hat{O}) G(y, x; t', x') = \delta(x-x') \delta(y-y').
\end{equation}
The relation between the propagator and the Green function is given by
\begin{equation}\label{eq36}
G(y, x; y', x') = \Theta(y-y') K(y, x; y', x') + A K(y, x; y', x'),
\end{equation}
where $\Theta(y - y')$ is the Heaviside step function and $A$ is a constant. 

\subsection{The Green function of Eq. (\ref{eq28})}

In this subsection we will find the Green function relevant to Eq. (\ref{eq28}) denoted by $G(T^{\alpha}, x; \tilde{T}^{\alpha}, \tilde{x})$. For that purpose we solve Eq. (\ref{eq35}) with $\hat{O} = - {^C\!\partial_{x}^{\alpha}}$ and $y = (t/\tau_0)^{\alpha} \equiv T^{\alpha}$ by taking twice the Laplace transform of this equation and applying \cite[Eq. (17.13.95) on p. 1115]{ISGradsteyn07}. The twofold use of Laplace transform, one in $T^{\alpha}$ ($T > 0$) and another one in $x$ ($x > 0$) gives
\begin{equation*}
\mathcal{L}\left[\mathcal{L}\left[(\partial_{T^{\alpha}} + {^C\!\partial_{x}^{\alpha}}) G(T^{\alpha}, x; \tilde{T}^{\alpha}, \tilde{x}); \tau\right]; \kappa\right] \!=\! e^{-\tau \tilde{T}^{\alpha}} e^{-\kappa \tilde{x}},
\end{equation*}
where $T > \tilde{T}$. The Laplace transform of the first derivative with respect to $T^{\alpha}$ is proportional to $\tau$, i.e. $\mathcal{L}[\partial_{T^{\alpha}} G(T^{\alpha}, x; \tilde{T}^{\alpha}, \tilde{x}); \tau] = \tau G_{L_{\tau}}(\tau, x; \tilde{T}^{\alpha}, \tilde{x})$. Similarly the Laplace transform of the fractional derivative in the Caputo sense of $G(T^{\alpha}, x; \tilde{T}^{\alpha}, \tilde{x})$ is given by $\kappa^{\alpha}G_{L_{\kappa}L_{\tau}}(\tau, \kappa; \tilde{T}^{\alpha}, \tilde{x}) - \kappa^{\alpha-1} G(T^{\alpha}, \tilde{x}; \tilde{T}^{\alpha}, \tilde{x})$, see \cite[Eq. (2.140) on p. 80 for $0 <\alpha<1$]{IPodlubny99} or \cite[Eq. (6) for $0 <\alpha<1$]{VKiryakova13}. The symbols $L_{\tau}$ and $L_{\kappa}$ denote the Laplace transforms with respect to time $T^{\alpha}$ and space $x$ such that $G_{L_{\kappa} L_{\tau}}(\tau, \kappa; t', x') = \mathcal{L}[\mathcal{L}[G(T^{\alpha}, x; \tilde{T}^{\alpha}, \tilde{x}); \tau]; \kappa]$. Assuming that $G(T^{\alpha}, \tilde{x}; \tilde{T}^{\alpha}, \tilde{x}) = 0$, we get
\begin{equation}\label{eq37}
G_{L_{\kappa} L_{\tau}}(\tau^{\alpha}, \kappa; \tilde{T}^{\alpha}, \tilde{x}) = \frac{e^{-\tau \tilde{T}^{\alpha}} e^{-\kappa \tilde{x}}}{\tau + \kappa^{\alpha}},
\end{equation}
Thus, the Green function $G(T^{\alpha}, x; \tilde{T}^{\alpha}, \tilde{x})$ can be obtained by taking twice the inverse Laplace transform of Eq. (\ref{eq37}):
\begin{align}\label{eq38}
& G(T^{\alpha}, x; \tilde{T}^{\alpha}, \tilde{x}) = \mathcal{L}^{-1}[\mathcal{L}^{-1}[G_{L_{\kappa} L_{\tau}}(\tau, \kappa; \tilde{T}^{\alpha}, \tilde{x}); T^{\alpha}]; x] \nonumber \\
& \qquad = \mathcal{L}^{-1}\left[\Theta(T^{\alpha} - \tilde{T}^{\alpha}) e^{-\kappa \tilde{x}} e^{-(T^{\alpha} - \tilde{T}^{\alpha}) \kappa^{\alpha}}; x\right] \nonumber \\
& \qquad = \Theta(x - \tilde{x}) \Theta(T^{\alpha} - \tilde{T}^{\alpha}) g_{\alpha}(T^{\alpha} - \tilde{T}^{\alpha}, x - \tilde{x}).
\end{align}
In Eq. (\ref{eq38}) we have  applied \cite[Eq. (3-3-11) on p. 146 of vol. 2]{INSneddon74} and the definition of the one-sided L\'{e}vy stable distribution. Comparing Eqs. (\ref{eq38}) and (\ref{eq36}) we have $A=0$ and 
\begin{equation*}
K(T^{\alpha}, x; \tilde{T}^{\alpha}, \tilde{x}) = \Theta(x - \tilde{x}) g_{\alpha}(T^{\alpha} - \tilde{T}^{\alpha}, x - \tilde{x})
\end{equation*}
with $g_{\alpha}(T^{\alpha} - \tilde{T}^{\alpha}; x - \tilde{x})$ given in Eq. (\ref{eq5}). From the properties of the one-sided L\'{e}vy stable distribution it can be seen that the propagator $K$ satisfies properties (i), (ii) and (iii) which are listed at the beginning of this Section. The property (ii) can be easily shown to hold by demonstrating that the Laplace transform $\mathcal{L}[(\partial_{T^{\alpha}} + {^C\!\partial^{\alpha}_x})\Theta(x - \tilde{x}) g_{\alpha}(T^{\alpha} - \tilde{T}^{\alpha}; x - \tilde{x}); \kappa]$ vanishes. It comes from \cite[Eq. (3-3-11) on p. 146]{INSneddon74} and \cite[Eq. (2.140) on p. 80]{IPodlubny99}. The property (iii) is proven in Appendix A.

\subsection{The Green function of Eq. (\ref{eq33})}

The Green function $\tilde{G}(T^{\alpha}, x; \tilde{T}^{\alpha}, \tilde{x})$ for $T > \tilde{T}$ which is the solution of Eq. (\ref{eq33}), i.e. Eq. (\ref{eq35}) with $\hat{O} = -\,{^R\partial_{|x|}^\alpha}$ and $y = (t/\tau)^{\alpha} \equiv T^{\alpha}$, can be obtained by taking the Laplace transform in $T^{\alpha}$ for $T > 0$ and Fourier transform in $x$ for $x\in\mathbb{R}$:
\begin{equation}\label{eq39}
\mathcal{F}\!\left[\mathcal{L}\!\left[(\partial_{T^{\alpha}} -\,{^R\partial_{|x|}^\alpha})\tilde{G}(T^{\alpha}\!, x; \tilde{T}^{\alpha}\!, \tilde{x}); \tau\right]\!\!; \kappa\right] = e^{-\tau \tilde{T}^{\alpha}}.
\end{equation}
The Fourier transform of the fractional derivative in the Riesz sense is equal to $-|\kappa|^{\alpha} \tilde{G}_{F_{\kappa}}(T^{\alpha}, \kappa; \tilde{T}^{\alpha}, \tilde{x})$, see \cite[Eq. (1.17)]{RGorenflo98} and \cite{SGSamko93}. Thus, Eq. (\ref{eq39}) reads 
\begin{equation*}
\tilde{G}_{F_{\kappa} L_{\tau}}(\tau, \kappa; \tilde{T}^{\alpha}, \tilde{x}) = \frac{e^{-\tau \tilde{T}^{\alpha}}}{\tau + |\kappa|^{\alpha}},
\end{equation*}
where $\tilde{G}_{F_{\kappa} L_{\tau}}(\tau, \kappa; \tilde{T}^{\alpha}, \tilde{x}) = \mathcal{F}[\mathcal{L}[\tilde{G}(T^{\alpha}, x; \tilde{T}^{\alpha}, \tilde{x}); \tau]; \kappa]$. After calculating the inverse Laplace transform and then the inverse Fourier transform of $\tilde{G}_{F_{\kappa} L_{\tau}}(\tau, \kappa; t', x')$, we get
\begin{align}\label{eq40}
\tilde{G}(T^{\alpha}, x; \tilde{T}^{\alpha}, \tilde{x}) & = \mathcal{F}^{-1}[\mathcal{L}^{-1}[\tilde{G}_{F_{\kappa} L_{\tau}}(\tau, \kappa; \tilde{T}^{\alpha}, \tilde{x}); T^{\alpha}]; x] \nonumber \\
& = \mathcal{F}^{-1}[\Theta(T^{\alpha} - \tilde{T}^{\alpha}) e^{-T^{\alpha} |\kappa|^{\alpha}}; x] \nonumber \\
& = \Theta(T^{\alpha} - \tilde{T}^{\alpha}) \tilde{g}_{\alpha}(T^{\alpha} - \tilde{T}^{\alpha}, x - \tilde{x}). 
\end{align}
The two-sided L\'{e}vy stable distribution $\tilde{g}_{\alpha}(T^{\alpha} - \tilde{T}^{\alpha}, x - \tilde{x})$ is defined in Eq. (\ref{eq11}). Comparison of Eq. (\ref{eq40}) with Eq. (\ref{eq36}) indicates that $A=0$ and propagator $\tilde{K}$ is given by
\begin{equation}\label{eq41}
\tilde{K}(T^{\alpha}, x; \tilde{T}^{\alpha}, \tilde{x}) =  \tilde{g}_{\alpha}(T^{\alpha} - \tilde{T}^{\alpha}, x - \tilde{x}).
\end{equation}
We stress that the properties of $\tilde{g}_{\alpha}(T^{\alpha}, x)$ imply that the requirements (i)-(iii) are satisfied for $\tilde{K}$ defined in Eq. (\ref{eq41}). For example, the property (ii)  follows from \cite[Eq. (2-3-8) on p. 39 of vol. 1]{INSneddon74} and \cite[Eq. (1.17)]{RGorenflo98}.

\section{The evolution operator method}\label{sec:5}
\setcounter{section}{5}
\setcounter{equation}{0} 

\subsection{The case of Eq. (\ref{eq28})}

From the mathematical point of view, Eq. (\ref{eq28}) with the initial condition $h(x)$ is the Cauchy-like problem. Its formal solution for $x > 0$ and $0 < \alpha < 1$ can be expressed via evolution operator $\hat{U}_{\alpha}(T)$ in the following way
\begin{align}\label{eq42}
& g_{\alpha}(T^{\alpha}, x) = \hat{U}_{\alpha}(T) h(x) = \exp(-T^{\alpha}\, {^C\!\partial_x^\alpha}) h(x) \\
& \qquad = \int_{0}^{\infty} g_{\alpha}(T^{\alpha}, \xi) h(x - \xi) d\xi, \label{eq43}
\end{align}
The second equality in Eq. (\ref{eq42}) has a symbolic meaning which suggests naively employing a series representation of $\hat{U}_{\alpha}(T)$. This series representation does not constitute an effective tool because the series obtained usually converges for fixed $T$ only and for special values of initial conditions. The use of integral representation, see Eq. (\ref{eq43}), for an established domain of $\hat{U}_{\alpha}(T)$ avoids the problem of convergence and opens the new way of finding and obtaining the mathematically correct solutions.

The integral kernel $g_{\alpha}(T^{\alpha}; x)$, given in Eqs. (\ref{eq6}) and (\ref{eq7}), is a special solution of Eq. (\ref{eq43}) for the initial condition $h(x) = \delta(x)$. It is also the propagator $K(T^{\alpha}, x; \tilde{T}^{\alpha}, \tilde{x})$ for $\tilde{T} = 0$ and $\tilde{x} = 0$. Eq. (\ref{eq43}) for $h(x) = g_{\alpha}(0; x)$ reconstructs the convolution property Eq. (\ref{eq13}) for $(t_2 - t_1)/\tau_0 = T$ and $(t_1 - t_0)/\tau_0 = 0$.  It can be also shown that for different times $t_0$, $t_1$ and $t_2$ such that $t_0 \leq t_{1} \leq t_{2}$, the evolution operator satisfies
\begin{equation}\label{eq44}
\hat{U}_{\alpha}(\ulamek{t_2 - t_0}{\tau_0}) = \hat{U}_{\alpha}(\ulamek{t_2 - t_1}{\tau_0}) \hat{U}_{\alpha}(\ulamek{t_1 - t_0}{\tau_0}).
\end{equation}
which follows from Eq. (\ref{eq13}).

Let us now establish the conditions under which Eq. (\ref{eq43}) solves Eq. (\ref{eq28}). For that purpose we calculate the Laplace transform of the both sides of Eq. (\ref{eq28}). Thus, we get
\begin{equation}\label{eq45}
\mathcal{L}[\partial_{T^{\alpha}} g_{\alpha}(T^{\alpha}, x); \kappa] = \partial_{T^{\alpha}} e^{- (T \kappa)^{\alpha}} = -\kappa^{\alpha} e^{- (T\kappa)^{\alpha}},
\end{equation}
where we applied Eq. (\ref{eq2}). The r.h.s of Eq. (\ref{eq28}), where we employed Eq. (\ref{eq43}) and \cite[Eq. (2.140) on p. 80 for $0 <\alpha<1$]{IPodlubny99}, reads
\begin{align}
\mathcal{L}[{^C\!\partial^{\alpha}_x} g_{\alpha}(T^{\alpha}, x); \kappa] & = \mathcal{L}\left[\int_0^\infty g_{\alpha}(T^{\alpha}, \xi) {^C\!\partial^{\alpha}_x} h(x-\xi) d\xi; \kappa\right] \nonumber \\
& = \int_0^\infty g_{\alpha}(T^{\alpha}, \xi) \mathcal{L}[{^C\!\partial^{\alpha}_x} h(x-\xi); \kappa] d\xi \nonumber \\
& = \int_0^\infty g_{\alpha}(T^{\alpha}, \xi) \kappa^\alpha \mathcal{L}[h(x-\xi); \kappa] \nonumber \\
& - \int_0^\infty g_{\alpha}(T^{\alpha}, \xi)d\xi\,  \kappa^{\alpha -1} h(0) \nonumber\\
& = \kappa^{\alpha} e^{-(T\kappa)^{\alpha}} - \kappa^{\alpha -1} h(0). \label{eq46}
\end{align}
If we assume that the Laplace transform of $h(x)$ exists, i.e. $\mathcal{L}[h(x); \kappa] = H(\kappa)$, $h(0) = 0$, and we change the sign like in Eq. (\ref{eq28}), then Eq. (\ref{eq46}) is equal to Eq. (\ref{eq45}). That implies that  $g_{\alpha}(t; x)$ given in Eq. (\ref{eq43}) is the solution of Eq. (\ref{eq28}).

The inverse Laplace transform of $h(x-\xi)$, which is equal to $e^{-\xi\kappa} \mathcal{L}^{-1}[H(\kappa); x]$, implies that the formal solution (\ref{eq43}) obtained by using the evolution operator method has also the form:
\begin{equation}\label{eq47}
g_{\alpha}(T^{\alpha}, x) = \mathcal{L}^{-1}[e^{- (T\kappa)^{\alpha}} H(\kappa); x],
\end{equation}
which allows one to rewrite $g_{\alpha}(T^{\alpha}, x)$ in terms of the inverse Mellin transform \cite{INSneddon74} 
\begin{equation*}
g_{\alpha}(T^{\alpha}, x) = \mathcal{M}^{-1}\left[\int_0^\infty \frac{\kappa^{-s} e^{- (T\kappa)^{\alpha}}}{\Gamma(1-s)} H(\kappa) d\kappa; x\right],
\end{equation*}
where we applied \cite[Eq. (2.1)]{JSLew75}.

\subsection{The case of Eq. (\ref{eq33})}

The formal solution obtained via the evolution operator method of Eq. (\ref{eq33}) with the initial condition $\tilde{h}(x)$, can be presented as
\begin{align}
\tilde{g}_{\alpha}(T^{\alpha}, x) & = \hat{V}_{\alpha}(T) \tilde{h}(x) =  e^{ T^{\alpha} {^R\!\partial^\alpha_{|x|}}} \tilde{h}(x) \nonumber \\
& = \int_{-\infty}^{\infty} \tilde{g}_{\alpha}(T^{\alpha}, \xi) \tilde{h}(x-\xi) d\xi. \label{eq48}
\end{align}
Similarly, like in the case of the operator $\hat{U}_{\alpha}(T)$, the integral representation (\ref{eq48}) frees us from using the series representation of $\hat{V}_{\alpha}(T)$ which, even for well-defined initial conditions, can lead to divergent solutions.

The integral kernel $\tilde{g}_{\alpha}(T^{\alpha}, \xi)$ is the two-sided L\'{e}vy stable distribution, see Eqs. (\ref{eq11}) and (\ref{eq12}), which is a special solution of Eq. (\ref{eq33}) with the initial condition $\tilde{h}(x) = \delta(x)$. It is the propagator $\tilde{K}(T^{\alpha}, x; \tilde{T}^{\alpha}, \tilde{x})$, see Eq. (\ref{eq41}), for $\tilde{T} = 0$ and $\tilde{x} = 0$. Eq. (\ref{eq48}) reproduces the convolution property Eq. (\ref{eq14}) for $\tilde{h}(x) = \tilde{g}_{\alpha}(0, x)$. In analogy to Eq. (\ref{eq44}), it can be shown that for $t_0 \leq t_1 \leq t_2$ the condition
\begin{equation*}
\hat{V}_{\alpha}(\ulamek{t_2-t_0}{\tau_0}) = \hat{V}_{\alpha}(\ulamek{t_2-t_1}{\tau_0}) \hat{V}_{\alpha}(\ulamek{t_1-t_0}{\tau_0})
\end{equation*}
is satisfied which stems form Eq. (\ref{eq14}).

Taking the Fourier transform of both sides of Eq. (\ref{eq33}) with $\tilde{g}_{\alpha}(T^{\alpha}, x)$ defined in Eq. (\ref{eq48}), we check that it is the correct solution of this differential equation. The Fourier transform of Eq. (\ref{eq33}) gives:
\begin{equation}\label{eq49}
\mathcal{F}[\partial_{T^{\alpha}} \tilde{g}_{\alpha}(T^{\alpha}, x); \kappa] = \partial_t e^{- T^{\alpha} |\kappa|^\alpha} = - |\kappa|^{\alpha} e^{- T^{\alpha} |\kappa|^\alpha},
\end{equation}
where we employed Eq. (\ref{eq11}). The Fourier transform of the r.h.s. of Eq. (\ref{eq33}) reads
\begin{align}
& \mathcal{F}[{^R\!\partial_{|x|}^{\alpha}} \tilde{g}_{\alpha}(T^{\alpha}, x); \kappa] = \mathcal{F}\left[\int_{-\infty}^{\infty} \tilde{g}_{\alpha}(T^{\alpha}, x) {^R\!\partial_{|x|}^{\alpha}} \tilde{h}(x-\xi) d\xi; \kappa\right] \nonumber \\
& \qquad = \int_{-\infty}^{\infty} \tilde{g}_{\alpha}(T^{\alpha}, x) \mathcal{F}[{^R\!\partial_{|x|}^{\alpha}} \tilde{h}(x-\xi); \kappa] d\xi \label{eq50} \\
& \qquad = - |\kappa|^{\alpha} \mathcal{F}\left[\int_{-\infty}^{\infty} \tilde{g}_{\alpha}(T^{\alpha}, x) \tilde{h}(x-\xi) d\xi; \kappa\right] \nonumber \\
& \qquad = - |\kappa|^{\alpha} \mathcal{F}\left[\tilde{g}_{\alpha}(T^{\alpha}, x); \kappa\right]. \nonumber
\end{align}
Using Eq. (\ref{eq8}) we obtain Eq. (\ref{eq49}) which completes the proof. In Eq. (\ref{eq50}) we employed \cite[Eq. (1.17)]{RGorenflo98} and assume that the Fourier transform of $\tilde{h}(x)$ exists and is equal to $\mathcal{F}[\tilde{h}(x); \kappa] = \tilde{H}(\kappa)$.

\section{Examples}\label{sec:6}
\setcounter{section}{6}
\setcounter{equation}{0}

\noindent
\textbf{(A)} We start with finding the formal solution of Eqs. (\ref{eq28}) and (\ref{eq33}) for the initial conditions given by the one- and two-sided L\'{e}vy stable distributions, $h(x) = g_{\beta}(0, x)$ and $\tilde{h}(x) = \tilde{g}_{\beta}(0, x)$, respectively. If the values of $\beta$ are the same as in the integral kernel of  Eqs. (\ref{eq43}) and (\ref{eq48}), i.e. $\beta = \alpha$, then we get the convolution properties Eqs. (\ref{eq13}) and (\ref{eq14}) for $(t_2 - t_1)/\tau_0 = T$ and $(t_1 - t_0)/\tau_0 = 0$. For $\beta \neq \alpha$ the formal solutions given in Eqs. (\ref{eq43}) and (\ref{eq48}) are presented in \cite{KGorska15}. \\

\noindent
\textbf{(B)} Eq. (\ref{eq43}) for the integral kernel $g_{1/2}(\sqrt{T}, \xi)$ given by the L\'{e}vy-Smirnov distribution and the initial condition $h(x) = (\pi x)^{-1/2} [1 - \exp(-\ulamek{1}{4x})]$ will be solved by using Eq. (\ref{eq47}). The Laplace transform of $h(x)$ is obtained by using \cite[Eq. (2.3.3.1) on p. 322 and Eq. (2.3.16.2) on p. 344]{APPrudnikov_v1} and is equal to $H(\kappa) = \kappa^{-1/2}(1 - e^{-\sqrt{\kappa}})$. Inserting it into Eq. (\ref{eq47}), employing for the Mellin transform $\mathcal{M}[f(x); s] = f_M(s)$ the formulas \cite[Eqs. (8.4.1.4), (8.4.1.7) and (8.4.1.5) on p. 531]{APPrudnikov_v3}, i.e. $\mathcal{M}^{-1}[a^s f_{M}(s); x] = \mathcal{M}^{-1}[f_{M}(s); \ulamek{x}{a}]$, $\mathcal{M}^{-1}[f_M(-s); x] = \mathcal{M}^{-1}[f_M(s); x^{-1}]$ and {$\mathcal{M}^{-1}[f_M(s+a); x] = x^{a}\mathcal{M}^{-1}[f_M(s); x]$, we get
\begin{align*}
g^{(B)}_{1/2}(\sqrt{T}, x) & = \frac{1}{\sqrt{\pi x}}\mathcal{M}^{-1}\left[\Gamma(s); \frac{T}{4 x}\right] \\
& - \frac{1}{\sqrt{\pi x}}\mathcal{M}^{-1}\left[\Gamma(s); \frac{(1+ \sqrt{T})^2}{4 x}\right].
\end{align*}
From \cite[Eq.(5.1) on p. 191]{FOberhettinger74} $g^{(B)}_{1/2}(\sqrt{T}, x)$ reads 
\begin{equation*}
g^{(B)}_{1/2}(\sqrt{T}, x) = \frac{1}{\sqrt{\pi x}}\left[e^{-\ulamek{T}{4x}} - e^{-\ulamek{(\sqrt{T}+1)^2}{4x}}\right],
\end{equation*}
see also Fig. \ref{fig2}, for different values of $T$. \\
\begin{figure}[!h]
\begin{center}
\includegraphics[scale=0.4]{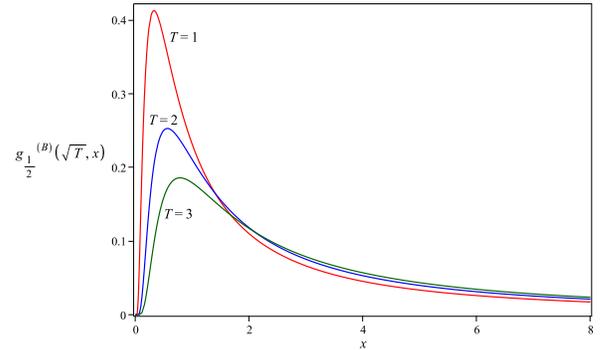}
\caption{\label{fig2} The plot of $g^{(B)}_{1/2}(\sqrt{T}, x)$ for $T=1$ (red curve), $T=2$ (blue curve), and $T=3$ (green curve).}
\end{center}
\end{figure}

\noindent
\textbf{(C)} The formal solution given in Eq. (\ref{eq48}) with the integral kernel being the Gauss distribution and the initial condition 
\begin{equation*}
h(x) = \frac{1}{b-a}\left\{\begin{array}{c c c} 0, & \text{for} & -\infty < x < a \\ 1, & \text{for} & a \leq x \leq b \\ 0, & \text{for} & b < x < \infty \end{array}\right.
\end{equation*}
can be calculated as follows
\begin{align}\label{eq51}
g^{(C)}_{2}(T^{2}, x) & = \int_{-\infty}^\infty \frac{1}{2 T \sqrt{\pi}} e^{-\xi^2/(4 T^2)} h(x - \xi) d\xi \nonumber \\
& = \frac{1}{2 T \sqrt{\pi}(b-a)} \int_{x-b}^{x-a} e^{-\xi^2/(4 T^2)} d\xi \\
& = \frac{1}{2(b-a)}\left[ {\rm erf}\left(\frac{b-x}{2 T}\right) - {\rm erf}\left(\frac{a-x}{2 T}\right)\right], \nonumber
\end{align}
and is plotted in Fig. \ref{fig3} for several values of $T$. In Eq. (\ref{eq51}) we have employed \cite[Eq. (2.33.2) on p. 108]{ISGradsteyn07}. 
\begin{figure}[!h]
\begin{center}
\includegraphics[scale=0.42]{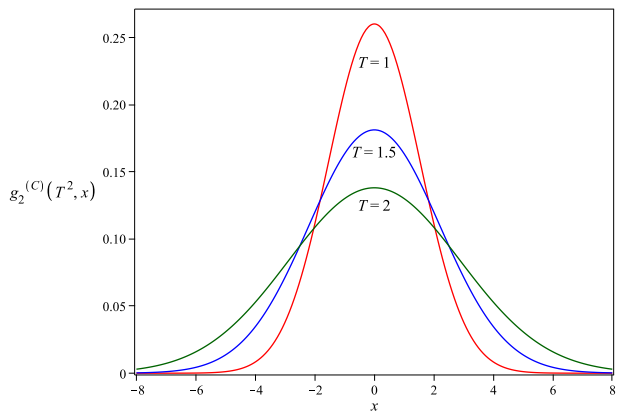}
\caption{\label{fig3} The plot of $g^{(C)}_{2}(T^{2}, x)$ for $a=-1$, $b=1$, and $T=1$ (red curve), $T=1.5$ (blue curve), and $T=2$ (green curve).}
\end{center}
\end{figure}

All the initial conditions chosen above are normalizable. Consequently, the evolution solutions of (A), (B) and (C) conserve the normalizability. We observed also that non-normalizable initial conditions, not considered here, lead to evolution solutions with diverging normalization.

\section{Conclusions}\label{sec:7}
\setcounter{section}{7}
\setcounter{equation}{0}

In this paper we have studied the consequence of employing the convolution properties of two-variable versions of the L\'{e}vy stable distributions where both arguments can vary. The natural evolution requirement imposed on the stretched or compressed functions allows us to derive new types of convolution properties of the L\'{e}vy stable distributions. They are, in fact, the evolution equations written in the integral form. Subsequently, using the tools of the complex analysis we derived the differential forms of these equations, namely the anomalous diffusion equations with the spacial fractional derivative in the Caputo or Riesz senses. The Caputo differential operator is obtained for $0 < \alpha < 1$ and $x>0$ and it is related to the one-sided L\'{e}vy stable distributions. The Riesz operator is obtained for $1 < \alpha < 2$ and $x\in\mathbb{R}$ and corresponds to the two-sided L\'{e}vy stable distributions \cite{KGorska11}. Our approach is similar to that of \cite{AISaichev97}, the difference being that out starting point is the physically motivated convolution of stretched exponentials, see Eqs. \eqref{eq13} and \eqref{eq14} above. This in turn leads to fractional differential equations and permits to write down their explicit solutions, see Eqs. \eqref{eq43} and \eqref{eq48}. We point out that the anomalous diffusion equations with the fractional derivative with respect to space coordinate describe the behavior of the random walker on a disordered space who makes the jumps of a length proportional to $x$ in the time interval from $T^{\alpha}$ to $T^{\alpha} + d T^{\alpha}$. 

We have also proposed the formal solution of two kinds of so obtained anomalous diffusion equations by employing the operational method. The integral representation of the appropriate evolution operators was found. Its integral kernels are the L\'{e}vy stable distributions which are given by the Green function connected with the anomalous diffusion equations. That allows us to define the self-reproducing solutions. We have also exemplified classes of the initial conditions for which our solutions work. Thus, we have shown that the formalism of evolution equation and the  integral transforms associated with them are very efficient tools to deal with evolution problems involving the space-fractional anomalous equations. For various fractional order of derivatives and different initial conditions we have presented some explicit examples of the solutions of space-fractional anomalous diffusion equations with two types of fractional derivatives operators. 

\section*{Acknowledgements}
\noindent
We thank the anonymous referees for constructive remarks. We thank Dr. {\L}. Bratek and Prof. K. Weron for important discussions. 

\noindent
K. G., A. H. and K. A. P. were supported by the PAN-CNRS program for French-Polish collaboration and the BGF scholarship founded by French Embassy in Warsaw, Poland. Moreover, K. G. thanks for support from MNiSW (Warsaw, Poland), "Iuventus Plus 2015-2016", program no IP2014 013073 and the Laboratoire d'Informatique de l'Universit\'{e} Paris-Nord in Villetaneuse (France) whose warm hospitality is greatly appreciated.

\appendix
\section{The derivations of Eqs. (\ref{eq13}) and (\ref{eq14})}\label{sec:8}
\setcounter{section}{8}
\setcounter{equation}{0} 

We start with the evolution property written for the stretched exponential function for $t_0 \leq t_1 \leq t_2$:
\begin{equation}\label{eq52}
e^{-\big(\ulamek{t_2-t_0}{\tau_0}\big)^{\alpha}} = e^{-\big(\ulamek{t_2-t_1}{\tau_0}\big)^{\alpha}} \circ e^{-\big(\ulamek{t_1-t_0}{\tau_0}\big)^{\alpha}}.
\end{equation}
The symbol '$\circ$' denotes a composition of functions which should be understood as the multiplication of stretched exponential functions defined in Eq. (\ref{eq3}). Thus, the r.h.s. of Eq. (\ref{eq52}) is equal to
\begin{align}\label{eq53}
&\text{RHS of Eq. (\ref{eq52})} = \!\!\int_0^{\infty}\!\! dy\, e^{-y} g_{\alpha}((\ulamek{t_2-t_1}{\tau_0})^{\alpha}, y) \nonumber\\
& \times \int_0^{\infty}\!\!\! dz\, e^{-z} g_{\alpha}((\ulamek{t_1-t_0}{\tau_0})^{\alpha}, z)\\ 
& = \!\!\int_0^{\infty}\!\! dy \!\!\int_{y}^{\infty}\!\!\! dx\, e^{-x} g_{\alpha}((\!\ulamek{t_2-t_1}{\tau_0}\!)^{\alpha}, y) g_{\alpha}((\!\ulamek{t_1-t_0}{\tau_0}\!)^{\alpha}, x-y) \nonumber\\
& = \!\!\int_0^{\infty}\!\!\! dx\, e^{-x} \!\!\left[\!\int_{0}^{x}\!\! dy\, g_{\alpha}((\!\ulamek{t_2-t_1}{\tau_0}\!)^{\alpha}, y) g_{\alpha}((\!\ulamek{t_1-t_0}{\tau_0}\!)^{\alpha}, x-y)\right]\!, \label{eq54}
\end{align}
where in Eq. (\ref{eq53}) the new variable $x = y + z$ is employed. After applying Eq. (\ref{eq3}) for the l.h.s. of Eq. (\ref{eq52}) and comparing with Eq. (\ref{eq54}) we obtain Eq. (\ref{eq13}).

To proof of Eq. (\ref{eq14}) begins with the evolution property for the compressed exponential:
\begin{equation}\label{eq55}
e^{-\big|\ulamek{t_2-t_0}{\tau_0}\big|^{\alpha}} = e^{-\big|\ulamek{t_2-t_1}{\tau_0}\big|^{\alpha}} \circ e^{-\big|\ulamek{t_1-t_0}{\tau_0}\big|^{\alpha}},
\end{equation}
where '$\circ$' denotes now the multiplication of two functions given in Eq. (\ref{eq9}). Substituting Eq. (\ref{eq9}) into Eq. (\ref{eq55}) and making similar changes of variable like in the case of Eq. (\ref{eq53}) we get Eq. (\ref{eq14}).

\section{Calculation of the second integral in Eq. (\ref{eq20})}\label{sec:9}
\setcounter{section}{9}
\setcounter{equation}{0}

The second integral in Eq. (\ref{eq20}) has the simple singularity at $y=x$, where $0 < x < \infty$. Taking the contour of integration which is the upper right one fourth of the circle, denoted by $Q_{R}$, with the pole at $y=x$, see Fig. \ref{fig1}, 
\begin{figure}[!h]
\begin{center}
\includegraphics[scale=0.3]{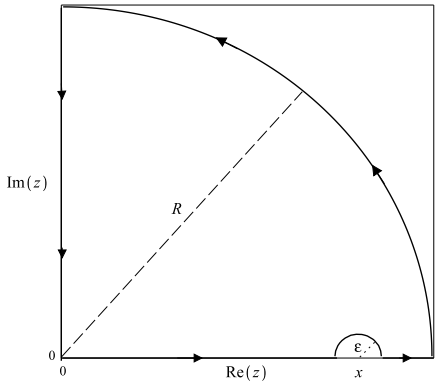}
\caption{\label{fig1} The contour of integration used in Eq. (\ref{eq56}).}
\end{center}
\end{figure}
and using the Cauchy's integral theorem \cite{FWByron92}, we find that 
\begin{align}\label{eq56}
& \int_0^x \frac{e^{-y\xi}}{x-y} dy = -\lim_{R\to\infty}\int_{Q_R} \frac{e^{-z\xi}}{x-z} dz - \lim_{\varepsilon\to0}\int_{S_{\varepsilon}} \frac{e^{-z\xi}}{x-z} dz \nonumber \\
& \quad - \lim_{R\to\infty}\int_{iR}^0 \frac{ie^{-ir\xi}}{x-ir} dr - \lim_{R\to\infty}\int_{x}^R \frac{e^{-y\xi}}{x-y} dy.
\end{align}

The imaginary part of the integral over the quadrant $Q_R$ vanishes. This can be shown by setting $z = Re^{i\theta}$ and studying the integral
\begin{equation}\label{eq57}
{\rm Im}\left[\lim_{R\to\infty}\int_{Q_R} \frac{e^{-z\xi}}{x-z} dz\right] = \lim_{R\to\infty}\int_0^{\pi/2} {\rm Im}[\psi(R e^{i\theta})] d\theta, 
\end{equation}
where $\psi(z) = i z e^{-\xi z}/(x-z)$ and ${\rm Im}[\psi(Re^{i\theta})]$ is an even function of $R$ and $\theta$ equals to 
\begin{align}\label{eq58}
{\rm Im}[\psi(Re^{i\theta})] & = \frac{R e^{-R\xi\cos\theta}}{x^2-2Rx \cos\theta+R^2} [(x\cos\theta-R) \nonumber\\
& \times \cos(R\xi\sin\theta) + x\sin\theta\sin(R\xi\sin\theta)].
\end{align}
Observe that Eq. (\ref{eq58}) for $\theta = \pi/2$ can be estimated by ${\rm Im}[\psi(i R)] \leq R(x-R)/(x^2 + R^2)$ which it is smaller or equal to ${\rm Im}[\psi(R)]$ for $0 < x < R$. That leads to 
\begin{equation*}
{\rm Im}[\psi(R e^{i \theta})] \leq {\rm Im}[\psi(R)] = \frac{Re^{-R\xi}}{x-R}.
\end{equation*}
After substituting it into Eq. (\ref{eq57}) it can be shown that the imaginary part of integral over the quadrant of radius $R$ goes to zero by choosing $R$ sufficiently large. 

We set $x - z = \varepsilon e^{i\phi}$ in the second integral in the right hand side of Eq. (\ref{eq56}), so that
\begin{align*}
\lim_{\varepsilon\to 0} \int_{S_{\varepsilon}} \frac{e^{-z\xi}}{x-z} dz & = - i e^{-x\xi } \lim_{\varepsilon\to0}  \int_0^\pi e^{\varepsilon \exp(i\phi)\xi} d\phi \nonumber \\
& = - i \pi e^{-x\xi}. 
\end{align*}
The third integral in the r.h.s. of Eq. (\ref{eq56}) after changing the variable of integration $ir = y$ and using \cite[Eq. (3.352.1) on p. 340]{ISGradsteyn07} gives the real function
\begin{equation*}
\int_{-R}^0 \frac{e^{-y\xi}}{x-y} dy = e^{-x \xi}\left[{\rm Ei}(R \xi + x \xi) - {\rm Ei}(x \xi)\right],
\end{equation*}
where ${\rm Ei}(x)$ is the exponential integral, see \cite[Section 8.2]{ISGradsteyn07}. In the forth integral in the r.h.s. of Eq. (\ref{eq56}) we change the variable $z$ as follows $x-z=u$. That gives
\begin{equation}\label{eq59}
\int_{x}^R \frac{e^{-y\xi}}{x-y} dy = e^{-x \xi} \int_{x-R}^0 \frac{e^{u \xi}}{u}.
\end{equation}
Using the series representation of the exponential it can be shown that Eq. (\ref{eq59}) is a real function which goes to infinity for $R\to\infty$. Concluding the considerations we find
\begin{equation}\label{eq60}
{\rm Im}\left[\int_0^x \frac{e^{-y\xi}}{x-y} dy \right] = \pi e^{-x\xi}.
\end{equation}

\end{document}